\documentstyle[twoside,cp]{article}

\newcommand{\beq}{\begin{equation}}
\newcommand{\eeq}{\end{equation}}
\newcommand{\beqa}{\begin{eqnarray}}
\newcommand{\eeqa}{\end{eqnarray}}
\newcommand{\bea}{\begin{eqnarray}}
\newcommand{\eea}{\end{eqnarray}}

\newcommand  {\BC}       {{\it Biophys.\ Chem.\ }}

\newcommand  {\EL}       {{\it Europhys.\ Lett.\ }}

\newcommand  {\JCP}      {{\it J.\ Chem.\ Phys.\ }}

\newcommand  {\JP}       {{\it J.\ Phys.\ }}

\newcommand  {\M}        {{\it Macromolecules\ }}

\newcommand  {\ProEng}   {{\it Protein\ Eng.\ }}

\newcommand  {\PNAS}     {{\it Proc.\ Natl.\ Acad.\ Sci.\ USA\ }}
\newcommand  {\PR}       {{\it Phys.\ Rev.\ }}
\newcommand  {\PRL}      {{\it Phys.\ Rev.\ Lett.\ }}

\input{psfig}

\begin{document}

\title{Folding and Design in Coarse-Grained Protein Models}

\author{Carsten Peterson \address{ 
Complex Systems, Department of Theoretical Physics, 
Lund University, S\"olvegatan 14A, \\
S-22362 Lund, Sweden; {\tt http://thep.lu.se/tf2/complex/} }}

\begin{abstract}
Recent advances in coarse-grained lattice and off-lattice protein 
models are reviewed.  The sequence dependence of thermodynamical 
folding properties are investigated and evidence for non-randomness 
of the binary sequences of good folders are discussed. Similar patterns 
for non-randomness are found for real proteins. Dynamical parameter 
MC methods, such as the tempering and multisequence algorithms, are 
essential in order to obtain these results. Also, a new MC method for design, 
the inverse of folding,  is presented. Here,  one maximizes conditional 
probabilities rather than minimizing energies. 
By construction, this method ensures that the designed sequences 
represent good folders thermodynamically. 

\end{abstract}

\maketitle

\section{Introduction}

Proteins are heterogenuous chain molecules composed of sequences 
of amino acids. The protein folding 
problem amounts to given a sequence of amino acids predict the 
protein 3D structure. There are 20 different amino acids. 
In the {\it Bioinformatics} approach one aims at extracting rules 
in a "black-box" manner by relating sequence with structure from 
databases. Here we pursue the physics approach, where given
interaction energies, the 3D structures and their thermodynamical 
properties are probed. In principle, this can be pursued on 
different levels of resolution. {\it Ab initio} quantum chemistry 
calculations can not handle the huge degrees of freedom, but are 
of course useful for estimating interatomic potentials. All-atom 
representations, where the atoms are the building blocks, also require 
very large computing resources for the full folding problem 
including thermodynamics, but are profitable for computing partial 
problems, binding energies etc. 

Here we pursue a course-grained representation, where the entities 
are the amino acids. This is motivated by the fact that the 
hydrophobic properties of the amino acids play a most 
important role in the folding process -- the amino acids that 
are hydrophobic (H) tend to form a core, whereas the hydrophilic or 
polar ones (P) are attracted to the surrounding H$_2$O solution.  
In such representations, the interactions between the amino acids 
and the solvent are reformulated in an {\it effective interaction} 
between the amino acids.

\section{Coarse-Grained Models}

Both lattice and off-lattice models have here been studied. 

A well studied lattice model is the HP model~\cite{Lau:89}
\beq
\label{HP}
E(r,\sigma) = 
-\sum_{i<j}\sigma_i \sigma_j\Delta(r_i - r_j)
\eeq
where $\Delta(r_i - r_j)=1$ if monomers $i$ and $j$ are 
non-bonded nearest neighbors and $0$ otherwise. For hydrophobic 
and polar monomers, one has $\sigma_i=1$ and 0, respectively. 
Being discrete, this model has the advantage that for sizes up to 
$N=18$ in 2D it can be solved exactly by exhaustive enumeration.

Similarly off-lattice models have been developed, where adjacent 
residues are linked by rigid bonds of unit length to 
form linear chains \cite{Irback:96b,Irback:96c}. The energy function 
is given by
\beq
E(r, \sigma)  = \sum_{i}F_i + 
\sum_{i<j}\epsilon(\sigma_i,\sigma_j)
[r_{ij}^{-12}-r_{ij}^{-6}] 
\label{energy}
\eeq
where $F_i$ is a local sequence-independent interaction chosen 
to mimic the observed local correlations among real proteins and 
the second term corresponds to amino-acid interactions, the 
strengths/signs of which are governed by 
$\epsilon(\sigma_i,\sigma_j)$.

\section{Folding}

Investigating thermodynamical properties of chains given by 
Eqs. (\ref{HP},\ref{energy}) is extremely tedious with standard 
MC methods; Metropolis, the hybrid method etc. Hence novel approaches are 
called for. {\it Dynamical Parameter} approaches have here 
turned out to be very powerful; the {\it tempering} 
\cite{Lyubartsev:92,Marinari:92} and {\it multisequence} \cite{Irback:95b} 
methods. In both approaches one enlarges the Gibbs distribution. In 
\cite{Lyubartsev:92,Marinari:92} one simulates 
\beq
P(r,k)= \frac{1}{Z} \exp(-g_k-E(r,\sigma)/T_k)
\eeq
with  ordinary $r$ and $k$ updates for 
$T_1<\ldots<T_K$, regularly quenching the system to the ground state. 
The weights are $g_k$ are chosen such 
that the probability of visiting the different $T_k$ is roughly 
constant. Similarly in the multisequence method the degrees of
freedom are enlarged to include different sequences according 
to
\beq
\label{joint}
P(r,\sigma)  =  \frac{1}{Z} \exp(-g_\sigma-E(r,\sigma)/T)\\
\eeq
where again $g_\sigma$ is a set of tunable parameters, 
which are subject to moves jointly with $r$. 

When estimating thermodynamical quantities, these dynamical 
parameter methods yield speedup factors of several orders 
of magnitude. 

A key issue when studying properties of protein models are 
to what extent different sequences yield structures with good 
folding properties from a thermodynamic standpoint. Defining 
good folding properties is straightforward in the lattice model case 
-- non-degenerate ground states. For off-lattice models a suitable 
measure can  be defined in terms of the mean-square
distance $\delta_{ab}^2$ between two arbitrary configurations $a$ 
and $b$. An informative measure of stability is the mean 
$\langle \delta^2\rangle$ \cite{Iori:91}. With a suitable cut on 
$\langle \delta^2\rangle$ good folders are singled out. For both 
lattice and off-lattice models, only a few \% of the sequences have good 
folding properties \footnote{Similar fractions are obtained within the 
replica approach for lattice models \cite{Garel/Shak}.}. 
When analyzing the sequence properties of good folders, 
one finds that similar signatures occur among real proteins when 
using a binary coding for the hydrophobicities \cite{Irback:96}. 
One might speculate that only those sequences with good folding 
properties survived the evolution.

\section{Design}

The ``inverse'' of protein folding, sequence optimization, is of 
utmost relevance in the context of drug design. Here, one aims at  
finding optimal amino acid sequences given a target structure such 
that the solution represents a good folder. This corresponds to maximizing 
the conditional probability \cite{Deutsch:9596},
\beqa
\label{P}
P(r_0|\sigma) = \frac{1}{Z(\sigma)}\exp (-E(r_0,\sigma)/T)\\
\label{Z}
Z(\sigma) ={\sum_r \exp(-E(r,\sigma)/T)}
\eeqa
Note that here $Z(\sigma)$ is not a constant quantity. A straightforward approach 
would therefore require a nested MC -- for each step in $\sigma$ a 
complete MC  has to be performed in $r$ \cite{Seno:96}. Needless to say, this 
is extremely time consuming. Various approximations for $Z$ has 
been suggested; chemical potentials fixing the net hydrophobicity 
and low-$T$ expansions \cite{Shakhnovich:93a}. Neither of these produce good 
folders in a reliable way.

\begin{figure}[tbh]
\vspace{-40mm}
\mbox{
  \hspace{-30mm}
  \psfig{figure=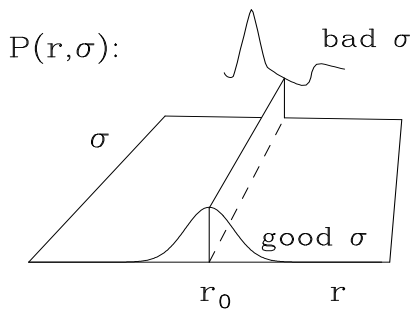,width=10cm,height=12cm}
}
\vspace{-50mm}
\caption{The distribution $P(r,\sigma)$. The choice of $g_\sigma$ 
(Eq.~(\ref{g})) implies that $P(r_0,\sigma)$ is flat in $\sigma$. 
Sequences not designing $r_0$ have maxima in $P(r_i | \sigma)$ for 
$r_i \not= r_0$ due to states with $E(r_i,\sigma)\leq E(r_0,\sigma)$. 
Sequence designing $r_0$ have unique maxima at $r=r_0$ in 
$P(r | \sigma)$, which for low $T$ contains most of the probability.}
\label{fig:1}
\end{figure}

Here we devise a different strategy based upon the 
multisequence method \cite{Irback:9799}. The starting point is the 
joint probability distribution (Eq. (\ref{joint}))
The corresponding marginal distribution is given by
\beqa
\label{marg}
P(\sigma)& = &\sum_r P(r,\sigma) = 
\frac{1}{Z} \exp(-g_\sigma) Z(\sigma) \nonumber \\
Z & = & \sum_\sigma \exp(-g_\sigma) Z(\sigma)  
\eeqa
With the choice 
\beq
g_\sigma=-E(r_0,\sigma)/T
\label{g}
\eeq
one obtains
\beq
P(r_0|\sigma)=\frac{P(r_0,\sigma)}{P(\sigma)}=\frac{1}{ZP(\sigma)}
\label{bayes}
\eeq
In other words, maximizing $P(r_0|\sigma)$ is in this case equivalent to 
minimizing $P(\sigma)$. This implies that bad sequences are visited 
more frequently than good ones in the simulation. This property 
may seem strange at a first glance. However, it can be used 
to eliminate bad sequences. The situation is illustrated in 
Fig.~\ref{fig:1}.  
Basically, one runs a MC in both $r$ and $\sigma$ using all 
(or a subset of) the sequences. Regularly, one estimates $P(\sigma)$. 
Sequences where $P(\sigma)$ exceeds a certain threshold are then 
eliminated, thereby purifying the sample towards designing 
sequences according to Eq. (\ref{bayes}). For lattice models one 
can use an alternative to eliminating high $P(\sigma)$ sequences, 
by removing sequences with $E(r,\sigma)\le E(r_0,\sigma)$.

Testing any design algorithm requires that one has access to 
{\it designable} structures, i.e. structures for which there 
exist good folding sequences. Furthermore, after the design 
process, it must be verified that the designed sequence indeed 
has the structure as a stable minimum (good folder). For $N \leq 18$ 
2D lattice models this is of course feasible, since these models 
can be enumerated exactly. For larger lattice models and off-lattice 
models this is not the case and testing the design approach  
is more laborious. 

Extensive tests have been performed for $N$=16, 18, 32 and 50 lattice 
and $N$=16 and 20 off-lattice chains respectively. For systems
exceeding $N$=20 one cannot go through all possible sequences. 
Hence a bootstrap procedure has been devised, where a set of 
preliminary runs with subsets of sequences is first performed. 
Positions along the chain with clear assignments of H or P are 
then clamped and the remaining degrees of freedom are run with all 
sequences visited. With no exceptions, the design algorithm efficiently 
singles out sequences that folds well into the (designable) target 
structures.

{\bf Acknowledgment:} The results reported here were obtained 
together with A. Irb\"ack, F. Potthast, E. Sandelin and O. Sommelius.

\end{document}